# 线性反馈系统中信息理论极限的分析

**摘要**：在线性反馈系统中，对于其控制性能的理论极限研究一直是近几十年来的热点问题之一。在本文中，利用信息理论的结果，对因果线性反馈系统进行了分析，得到了一个新的类似于 bode 极限形式基本性能不等式。这个不等式联系了有向信息，互信息和 bode 敏感度方程。包括了已有的关于反馈系统性能极限的结论。

# Analysis of Information Theoretic Limitation for Linear Time Invariant Feedback Systems.

*Abstract-* *Information-theoretic fundamental limitation in feedback control system is an important topic for decades. In this paper, a new bode-like fundamental inequality in causal feedback control system is developed. This inequality relates directed information, mutual information and bode sensitivity functions. This inequality recovers previous known results in feedback system as special cases.*

## I. 引言

对于反馈控制系统的性能极限研究是一直以来学术界的热点问题。因为反馈控制系统已经广泛的应用于工业界的各个领域，所以其性能极限的掌握显得尤为重要，并且在系统设计的时候也起到了指导性作用。从上个世纪 20 年代开始，控制领域的学者就开始用控制理论对反馈系统进行研究，并且取得了丰硕的成果[1]。

信息理论是通信领域的基础理论。从 shannon 开创了信息理论开始，就一直致力于研究开环通信系统的通信信道容量和可以达到信道容量的编码方法[2][3]。知道上个世纪九十年代，有向信息概念的提出使信息论开始应用于因果反馈系统的研究。有向信息是传统互信息的在因果系统中的扩展，并且可以用传统的熵来量化[4][5][6]。近些年来，人们发现有向信息可以用来度量带有无噪声反馈通信信道的信道容量[10]。从本质来讲，无噪声反馈通信系统在某种程度上等价于控制领域中的反馈控制系统。因此，有向信息作为一个桥梁，链接了信息论和反馈控制理论。并且通过信息论的结果，可以更好的理解和研究反馈控制系统中的性能极限[7-9][11-12] [15-18]。同时，控制理论中的方法和结论也可以用来解决信息论里的理论问题。

在本文中，基于信息论中的概念的结果，得到了一个新的反馈控制系统中性能极限的不等式。这个不等式对于系统设计和分析有着重要的意义。

## II. 信息论模型与概念

在本节中，考虑带有随机通信信道的反馈控制系统，将探讨一个关于反馈控制性能和信息率的基本理论限制条件。由于 Bode 积分极限原理在反馈控制系统中的重要性，之前的大量工作把这个积分极限在反馈控制问题里一直延伸到更普遍的框架。本节的理论结果基于渐进稳定性假设，可以看作是另一个 bode 积分极限的扩展结果。

首先，可逆通信信道可以定义如下。

**定义**. 如果一个随机通信信道的随机因子可以通过信道当前的输入和输出值得以确定，则这个随机信道是可逆通信信道。

例如，考虑一个加载高斯白噪声的通信信道
$$Y_i = X_i + W_i$$
噪声 W 是可以由信道的输入输出值（X，Y）得到，所以这个随机信道是可逆的。

接下来，定义信息速率如下。

**定义** 令 X 和 Y 是两个稳定随机过程，它们之间的信息速率定义为

$$I_\infty(X;Y) = \lim_{n->\infty} \frac{I(X_1^n;Y_1^n)}{n}$$

信息速率表明了在任意的通信信道里，可以以多快的速度可靠的传输信息。

在本文中，有向信息速率是很重要的概念。有向信息在信息论领域有着广泛的应用。它可以用来量化有理想反馈的通信信道的信道容量。对于有噪声反馈的通信信道，有向信息依然可以用来分析信道里的信息流，并且可以用来得到一些有噪声反馈通信信道容量的上界和下界[13-14] [20] [22]。除此之外，有向信息与经济学里的 Granger Causality，数据压缩等都有很直接的联系[21]，并且可以应用到目标跟踪[19]。有向信息和有向信息速率速率定义如下。

**定义**：令 X 和 Y 是两个稳定随机过程，它们之间的有向信息和有向信息速率定义为

$$I(X_1^n -> Y_1^n) = \sum_{i=1}^{n} I(Y_i; X_1^i | Y_1^{i-1})$$

$$I_\infty(X -> Y) = \lim_{n->\infty} \frac{I(X_1^n -> Y_1^n)}{n}$$

通过数学表达式可以看出，有向信息度量了信息流的因果关系。也就是说，序列 Y 是由序列 X 以时间顺序依次产生的。那么从序列 X 到序列 Y 的有向信息表明了从序列 Y 以因果顺序依次含有序列 X 的信息的总和。

接下来，本文引入功率谱密度和反馈控制理论中的敏感度方程。

**定义**：对于一个渐进稳定的随机信号 a，它的功率谱密度表示为 $S_a(e^{jw})$。

功率谱密度定义了信号或者时间序列的功率如何随频率分布。这里功率可能是实际物理上的功率，或者更经常便于表示抽象的信号被定义为信号数值的平方，也就是当信号的负载为 1 欧姆(ohm)时的实际功率。

**定义**：对于给定的两个渐进稳定的随机信号 a 和 b，b 相对于 a 的敏感度方程定义为

$$S_{a,b}(e^{jw}) = \sqrt{\frac{S_a(e^{jw})}{S_b(e^{jw})}}$$

这个敏感度方程并不等于传统反馈控制系统中的敏感度方程。它是传统敏感度方程的一个扩展。

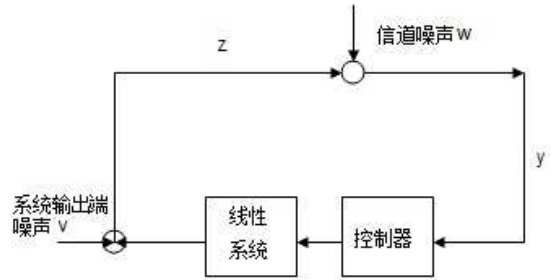

图 1. 含反馈噪声的线性反馈系统

### III. 反馈线性系统的性能极限不等式

接下来，基于信息理论的概念和反馈控制理论，推导出一个关于反馈系统中抵抗信道噪声能力的性能极限。

**定理**：假设一个反馈控制系统如图 1 所示。假设 P 是线性时不变系统，它的初始状态变量表示为 $X_0$。K 是控制器。信号 v 是系统输出端的加载噪声。信号 w 是反馈过程中加载的稳定高斯噪声。则下面的不等式是恒成立的，

$$I_\infty((X_0,V);Y) \leq \frac{1}{2\pi}\int_{-\pi}^{\pi} \log(S_{Y,W}(e^{jw}))dw$$

**证明**：首先，基于互信息的熵展开，

$$I(X_0;Y_1^n) = h(Y_1^n) - h(Y_1^n | X_0)。$$

利用熵展开的链式法则，

$$h(Y_1^n | X_0) = \sum_{i=1}^{n} h(Y_i | Y^{i-1}, X_0)$$

作进一步的展开，

$$\sum_{i=1}^{n} h(Y_i | Y^{i-1}, X_0) = \sum_{i=1}^{n} h(Y_i | Y^{i-1}, X_0, V^n)$$
$$+ \sum_{i=1}^{n} I(Y_i, V^n | Y^{i-1}, X_0)$$

利用互信息的链式法则，

$$\sum_{i=1}^{n} h(Y_i | Y^{i-1}, X_0) = \sum_{i=1}^{n} h(Y_i | Y^{i-1}, X_0, V^n)$$
$$+ I(Y^n; V^n | X_0)$$

对于熵，增加给定条件会减少熵的值，所以

$$\sum_{i=1}^{n} h(Y_i | Y^{i-1}, X_0) \leq \sum_{i=1}^{n} h(Y_i | Y^{i-1}, X_0, V^n, Z^i)$$
$$+ I(Y^n; V^n | X_0)$$

因为反馈的通信信道是可逆的，也就是说，

$$Y_i = Z_i + W_i$$

所以，上面的不等式进一步变为

$$\sum_{i=1}^{n} h(Y_i | Y^{i-1}, X_0) \leq \sum_{i=1}^{n} h(W_i | Y^{i-1}, X_0, V^n, Z^i)$$
$$+ I(Y^n; V^n | X_0)$$

$$\sum_{i=1}^{n} h(Y_i | Y^{i-1}, X_0) \leq \sum_{i=1}^{n} h(W_i | W^{i-1})$$
$$+ I(Y^n; V^n | X_0)$$

再次利用熵的链式法则，

$$\sum_{i=1}^{n} h(Y_i | Y^{i-1}, X_0) \leq h(W^n) + I(Y^n; V^n | X_0)$$

综合上面的推导，下面的不等式是成立的。

$$I((X_0, V^n); Y_1^n) = I(X_0; Y_1^n) + I(V^n; Y^n | X_0)$$
$$\leq h(Y^n) - h(W^n)$$

接下来，基于有向信息的定义，

$$I(Z^n -> Y^n) = \sum_{i=1}^{n} I(Y_i, Z^n | Y^{i-1})$$

再次利用互信息的熵展开，

$$I(Y_i, Z^n | Y^{i-1}) = h(Y_i | Y^{i-1}) - h(Y_i | Y^{i-1}, Z^i)$$

$$I(Y_i, Z^n | Y^{i-1}) = h(Y_i | Y^{i-1}) - h(W_i | Y^{i-1}, Z^i)$$
$$I(Y_i, Z^n | Y^{i-1}) = h(Y_i | Y^{i-1}) - h(W_i | W^{i-1}, Z^i)$$
$$I(Y_i, Z^n | Y^{i-1}) = h(Y_i | Y^{i-1}) - h(W_i | W^{i-1})$$

综合上面的推导可以得出，

$$I(Z^n -> Y^n) = h(Y^n) - h(W^n)$$

接下来，基于线性系统的假设，信号 y 的功率谱密度可以作为它的熵值得一个上界。具体可以表达为下面的不等式。

$$h_\infty(Y) \leq \frac{1}{2\pi} \int_{-\pi}^{\pi} \frac{1}{2} \log(2\pi e S_Y(e^{jw})) dw$$

如果信号 y 是稳定的高斯信号，则上面的不等式可以恒取等式。

同样，对于反馈信道中的噪声 w，它的熵值可以用功率谱密度来表达。

$$h_\infty(W) \leq \frac{1}{2\pi} \int_{-\pi}^{\pi} \frac{1}{2} \log(2\pi e S_W(e^{jw})) dw$$

有了以上的两个功率谱密度表达式，有向信息可以表达为功率谱密度的形式如下。

$$I_\infty(Z -> Y) = h_\infty(Y) - h_\infty(W)$$
$$\leq \frac{1}{2\pi} \int_{-\pi}^{\pi} \frac{1}{2} \log(2\pi e S_Y(e^{jw})) dw$$
$$- \frac{1}{2\pi} \int_{-\pi}^{\pi} \frac{1}{2} \log(2\pi e S_W(e^{jw})) dw$$
$$= \frac{1}{2\pi} \int_{-\pi}^{\pi} \frac{1}{2} \log(\frac{S_Y(e^{jw})}{S_W(e^{jw})}) dw$$
$$= \frac{1}{2\pi} \int_{-\pi}^{\pi} \log(S_{Y,W}(e^{jw})) dw$$

因为已知

$$I((X_0, V^n); Y_1^n) \leq h(Y^n) - h(W^n) = I(Z^n -> Y^n)$$

定理中的不等式

$$I_\infty((X_0, V); Y) \leq \frac{1}{2\pi} \int_{-\pi}^{\pi} \log(S_{Y,W}(e^{jw})) dw$$

是成立的。

这个定理融合了信息论中的互信息和反馈控制系统中的敏感度方程。互信息的值提供了反馈系统中抵抗信道噪声的一个下界。从这个定理可以看出，如果系统输出端的噪声 V 比较大，则互信息的值会变大。这就说明系统抵抗信道噪声的能力就要减弱。[23-24]对反馈系统中抵抗信道噪声的能力也做了相关研究。本文的结论可以看做是之前结论的补充和扩展。

## IV. 结束语

对于线性反馈系统的性能理论极限的研究已经有几十年的历史。但基于近年来信息理论的不断发展和完善，更多的信息理论的成果可以用来研究反馈系统的理论极限，是其极限的研究更加成果丰硕。本文利用了信息理论中的熵，互信息和有向信息的概念，推理出一个新的性能极限不等式。这个不等式建立了反馈系统中互信息，有向信息和 bode 敏感度方程的内在理论联系。本文的结论对以后的理论极限研究做了铺垫，基于本文的结论，更多的极限不等式可以被发现。这些将是未来的研究方向。

## 参考文献

[1] G. C. Goodwin, S. Graebe, and M. E. Salgado, Control System Design. New Jersey: Prentice Hall, 2001.

[2] T. M. Cover and J. A. Thomas, Elements of Information Theory, 2nd ed. Hoboken, N.J: Wiley-Interscience, 2006.

[3] R. W. Yeung, A first course in information theory. Springer, 2002..

[4] J. Massey, "Causality, feedback and directed information," in Proc. Intl. Symp. Inf. Theory and its Appl., Hawaii, USA, Nov. 1990, pp. 303–305.

[5] J. Massey and P. Massey, "Conservation of mutual and directed information," in Proc. IEEE Int. Symp. Information Theory, Sept. 2005, pp. 157–158

[6] G. Kramer, "Directed information for channels with feedback." Ph.D. dissertation, Swiss federal institute of technology, 1998.

[7] J. Baillieul, "Feedback Designs in Information Based Control," Stochastic Theory and Control: Proceedings of a Workshop Held in Lawrence, Kansas, 2002.

[8] A. Savkin, "Analysis and synthesis of networked control systems: Topological entropy, observability, robustness and optimal control," Automatica, vol. 42, pp. 51–62, 2006.

[9] G. Nair and R. Evans, "Stabilizability of stochastic linear systems with finite feedback data rates," SIAM Journal on Control and Optimization, vol. 43, no. 2, pp. 413–436, 2004.

[10] S. Tatikonda and S. Mitter, "The capacity of channels with feedback," IEEE Trans. Inf. Theory, vol. 55, no. 1, pp. 323–349, Jan. 2009.

[11] S. Tatikonda, A. Sahai, and S. Mitter, "Stochastic linear control over a communication channel," IEEE Transactions on Automatic Control, vol. 49, no. 9, pp. 1549–1561, 2004.

[12] S. Tatikonda and S. Mitter, "Control under communication constraints," IEEE Transactions on Automatic Control, vol. 49, no. 7, pp. 1056–1068, July 2004.

[13] C. Li, and N. Elia. "Bounds on the achievable rate of noisy feedback gaussian channels under linear feedback coding scheme." In Proc. IEEE Int. Symp. Information Theory, 2011, pp. 169-173

[14] C. Li, and N. Elia. "The information theorectic characterization of the capacity of channels with noisy feedback." In Proc. IEEE Int. Symp. Information Theory, 2011. pp. 174-178.

[15] N. Martins and M. Dahleh, "Feedback control in the presence of noisy channels: "Bode-Like'' fundamental limitations of performance," IEEE Trans. Autom. Control, vol. 53, no. 7, pp. 1604–1615, Aug. 2008.

[16] N. Martins, M. Dahleh, and J. Doyle, "Fundamental limitations of disturbance attenuation in the presence of side information," IEEE Transactions on Automatic Control, vol. 52, no. 1, pp. 56–66, 2007.

[17] N. Martins, M. Dahleh, and N. Elia, "Feedback stabilization of uncertain systems in the presence of a direct link," IEEE Transactions on Automatic Control, vol. 51, no. 3, pp. 438–447, 2006.

[18] N. Elia, "When Bode meets Shannon: Control oriented feedback communication schemes," IEEE Transactions on Automatic Control, vol. 49, no. 9, pp. 1477–1488, 2004

[19] H. Zhang and Y.-X. Sun, "Directed information and mutual information in linear feedback tracking systems," in Proc. 6-th World Congress on Intelligent Control and Automation, June 2006, pp. 723–727

[20] C. Li, and N. Elia. "Upper bound on the capacity of Gaussian channels with noisy feedback." In Communication, Control, and Coputing (Allerton), 49th Annual Allerton Conference on, 2011, pp. 84-89.

[21] H. H. Permuter, Y.-H. Kim, and T. Weissman, "Interpretations of directed information in portfolio theory, data compression, and hypothesis testing," IEEE Trans. Inf. Theory, vol. 57, no. 6, pp. 3248–3259, June 2011.

[22] C. Li, and N. Elia. "The information flow and capacity of channels with noisy feedback." arXiv preprint. arXiv: 1108.2815 (2011)

[23] H. Shingin and Y. Ohta, "Disturbance rejection with information constraints: Performance limitations of a scalar system for bounded and Gaussian disturbances," Automatica, vol. 48, pp. 1111–1116, 2012.

[24] J. Freudenberg, R. Middleton, and J. Braslavsky, "Stabilization with disturbance attenuation over a Gaussian channel," in Proceedings of the 46th IEEE Conference on Decision and Control, New Orleans, USA, 2007